\documentstyle[psfig]{mn}

\newif\ifAMStwofonts

\ifoldfss

  \ifCUPmtlplainloaded \else
    \NewTextAlphabet{textbfit} {cmbxti10} {}
    \NewTextAlphabet{textbfss} {cmssbx10} {}
    \NewMathAlphabet{mathbfit} {cmbxti10} {} 
    \NewMathAlphabet{mathbfss} {cmssbx10} {} 
  \fi
  \ifAMStwofonts
    \ifCUPmtlplainloaded \else
      \NewSymbolFont{upmath} {eurm10}
      \NewSymbolFont{AMSa} {msam10}
      \NewMathSymbol{\upi}     {0}{upmath}{19}
      \NewMathSymbol{\umu}     {0}{upmath}{16}
      \NewMathSymbol{\upartial}{0}{upmath}{40}
      \NewMathSymbol{\leqslant}{3}{AMSa}{36}
      \NewMathSymbol{\geqslant}{3}{AMSa}{3E}

      \let\leq=\leqslant \let\le=\leqslant
      \let\geq=\geqslant 
    \fi
  \fi
\fi 
\ifnfssone
  \newmathalphabet{\mathit}
  \addtoversion{normal}{\mathit}{cmr}{m}{it}
  \addtoversion{bold}{\mathit}{cmr}{bx}{it}

  \newmathalphabet{\mathbfit} 
  \addtoversion{normal}{\mathbfit}{cmr}{bx}{it}
  \addtoversion{bold}{\mathbfit}{cmr}{bx}{it}
  \newmathalphabet{\mathbfss} 
  \addtoversion{normal}{\mathbfss}{cmss}{bx}{n}
  \addtoversion{bold}{\mathbfss}{cmss}{bx}{n}
\ifAMStwofonts
    \ifCUPmtlplainloaded \else
      \UseAMStwoboldmath
      \makeatletter
      \new@mathgroup\upmath@group
      \define@mathgroup\mv@normal\upmath@group{eur}{m}{n}
      \define@mathgroup\mv@bold\upmath@group{eur}{b}{n}
      \edef\UPM{\hexnumber\upmath@group}
      \new@mathgroup\amsa@group
      \define@mathgroup\mv@normal\amsa@group{msa}{m}{n}
      \define@mathgroup\mv@bold\amsa@group{msa}{m}{n}
      \edef\AMSa{\hexnumber\amsa@group}
      \makeatother
      \mathchardef\upi="0\UPM19
      \mathchardef\umu="0\UPM16
      \mathchardef\upartial="0\UPM40
      \mathchardef\leqslant="3\AMSa36
      \mathchardef\geqslant="3\AMSa3E

      \let\leq=\leqslant \let\le=\leqslant
      \let\geq=\geqslant 
    \fi
  \fi
\fi 
\ifnfsstwo

  \DeclareMathAlphabet{\mathbfit}{OT1}{cmr}{bx}{it}
  \SetMathAlphabet\mathbfit{bold}{OT1}{cmr}{bx}{it}
  \DeclareMathAlphabet{\mathbfss}{OT1}{cmss}{bx}{n}
  \SetMathAlphabet\mathbfss{bold}{OT1}{cmss}{bx}{n}
  \ifAMStwofonts
    \ifCUPmtlplainloaded \else
      \DeclareSymbolFont{UPM}{U}{eur}{m}{n}
      \SetSymbolFont{UPM}{bold}{U}{eur}{b}{n}
      \DeclareSymbolFont{AMSa}{U}{msa}{m}{n}
      \DeclareMathSymbol{\upi}{0}{UPM}{"19}
      \DeclareMathSymbol{\umu}{0}{UPM}{"16}
      \DeclareMathSymbol{\upartial}{0}{UPM}{"40}
¤      \DeclareMathSymbol{\leqslant}{3}{AMSa}{"36}
      \DeclareMathSymbol{\geqslant}{3}{AMSa}{"3E}

      \let\leq=\leqslant \let\le=\leqslant
      \let\geq=\geqslant 
    \fi
  \fi
\fi 
\ifCUPmtlplainloaded \else
  \ifAMStwofonts \else 
    \def\upi{\pi}
    \def\umu{\mu}
    \def\upartial{\partial}
  \fi
\fi

\title{Dwarf Elliptical Galaxies:\\ Structure, 
         Star Formation, and  Color-Magnitude Diagrams} 

\author[G. Carraro et al.]
{Giovanni Carraro$^1$, Cesare Chiosi$^1$, L\'eo Girardi$^1$ and 
 Cesario Lia$^2$ \\
$^1$Dipartimento di Astronomia, Universit\'a di Padova, 
Vicolo Osservatorio 2,	I-35122, Padova, Italy\\
$^2$SISSA/ISAS, Via Beirut 2, I-34013, Trieste, Italy}

\date{}

\pagerange{\pageref{firstpage}--\pageref{lastpage}}
\pubyear{2001}

\begin{document}

\maketitle

\label{firstpage}

\begin{abstract} 
The aim of this paper is to cast light on the formation and evolution of 
elliptical galaxies by means of   N-body/hydro-dynamical simulations
that include star formation, feed-back and chemical evolution. Particular
attention is paid to the case of dwarf spheroidals of the 
Local Group   which, thanks to their proximity and modern ground-based and
space instrumentation, can be resolved into single 
stars so that independent determinations of their age and star formation
history can be derived. Indeed, the analysis of the 
color-magnitude diagram of their stellar content allows us  to infer
the past history of 
star formation and chemical enrichment thus setting important constraints 
on galactic models. 
Dwarf galaxies are known to exhibit complicated histories of star formation 
ranging from a single very old episode to a series of bursts  over
most of the Hubble time. By understanding the physical process driving 
star formation in these objects, we might  be able to infer the mechanism
governing star formation in more massive elliptical galaxies. Given these
premises, we start
 from 
virialized haloes of dark matter, and follow the infall of gas into the
potential wells and the formation of stars. We find that in objects of the same 
total mass, different star formation histories are possible, if the collapse 
phase started at different initial densities. 
We predict
the final structure of dwarf spheroidal galaxies,   their kinematics, their  
large scale distribution of gas and stars, and 
their detailed  histories of the star formation   and metal
enrichment. Using  a  population synthesis technique, star formation and
metal enrichment rates are then adopted to
generate  the present color-magnitude diagrams  of the  
stellar populations hosted by dwarf spheroidal galaxies. The simulations 
are made assuming the red-shift of galaxy formation $z_{for}=5$ and
varying the cosmological parameters $H_0$ and $q_0$.  
The resulting color-magnitude diagrams are then 
compared with the observational ones for some dwarf spheroidals 
of  the Local Group.
 
\end{abstract}

\begin{keywords}
Galaxy Formation - N-body Simulations - Stellar Populations - CM-Diagrams
\end{keywords}

\section{Introduction}

The present-day challenge with elliptical galaxies (EG's) is to unravel  
their formation and  evolutionary history. In a simplified picture of the 
issue, the problem can be cast
as follows. Do EG's form by hierarchical merging of pre-existing 
sub-structures (maybe disc galaxies) 
made of stars and gas? Was each merger  accompanied by strong
star formation?  Or conversely, do they originate  from the early aggregation
of lumps of gas turned into stars in the remote past via a 
burst-like  episode ever since followed by quiescence so as to mimic a 
sort of monolithic process? 
If so, what is the key physical parameter governing this happening?
Why do faint dwarf spheroidal and elliptical galaxies  show evidence of 
recent star formation, whereas the luminous elliptical galaxies  show
properties that are more compatible with very old stellar activity? 

The answers to the above questions may perhaps come from understanding the 
physical parameters governing the star formation history (SFH) of 
Dwarf Galaxies (DGs) of the Local Group.

These objects in fact are thought to be the most common type
of galaxies in the Universe. Therefore,  their stellar populations,  
their gas and dark matter (DM) content, their dynamical properties etc.
are the subject of a great deal of observational and theoretical studies 
(see Mateo 1998 for a recent review on the subject) to unravel the mechanism
of formation.

From the theoretical side, much effort has been spent to infer the star 
formation history of
DGs from their observed CM-diagrams (CMD), and to this purpose different 
sophisticated techniques have been developed
(Gallart et al. 1999, Hernandez et al. 2000 and references therein).

Re-constructing the star formation 
 history of these galaxies is of paramount importance 
in order to understand whether they formed in isolation or suffered from mergers
or tidal stripping during their evolution. 
This  galaxy types are indeed 
believed to be the building blocks for the assembly of larger galaxies.

Recent investigations proved that in many
cases the star formation history deduced from the CMD 
is very irregular. 
Nice examples are  the dwarf spheroidal (DSph) galaxies 
Carina and Leo\,I which together with a very old population
show the signature of recent star forming activity.

Other galaxies, however, seem to have
experienced a single burst of star formation, 
although rather extended in time, like
Draco, Ursa Minor or Sculptor, whose CMDs resemble those of 
globular clusters
in the Milky Way halo.

It is often argued that DSGs were born almost
at the same time, but evolved along different paths, probably depending
on their distance to the Milky Way, their orbit and the amount of interaction
experienced in the past (see Mateo 1998). Is this the sole possibility?

A sample of DSph galaxies is presented in Table~\ref{tab_data}, together with
a compilation of their basic properties. We show the CMD of two of them,
Carina and Sculptor, in Fig.~\ref{cmd_ca_scu}.  They represent
two very different evolutionary paths: in the case of Carina,
star formation seems to have lasted more than about 6 Gyr, where in the case of 
Sculptor
a dominant old burst of short duration seems to be more appropriate.

In this paper, instead of analyzing CMDs,
we propose a completely different approach, i.e. we intend to derive
from {\it first principles} the star formation history of dwarf galaxies
together with its large range of types.
Starting from "realistic" conditions for cosmological 
DM halos, we follow by means of N-body Tree-SPH simulations
the formation of dwarf galaxies out of the cooling gas
in the DM halo potential well,   and the transformation of gas into   stars.

The present analysis stems from the recent study by Carraro et al.
(2001), in which monolithic models of elliptical galaxies of different mass
have been investigated by means of similar N-body Tree-SPH models with
complete description of star formation, chemical enrichment, and energy 
feed-back. 
The key result of those simulations   is that 
starting from virialized haloes of DM
with the same initial density, 
at decreasing galaxy mass the star formation history 
gradually changes from a single sharply peaked initial episode
to a series of burst stretching over most of the Hubble time.
This is a very promising result because it considerably alleviates  several
points of weakness of SDGW model of Larson (1974) currently in use to 
interpret the  data of elliptical galaxies. 

Applying the Carraro et al. (2001) scheme to the family of dwarf galaxies
in which a large variety of star formation histories are seen for objects
spanning a rather small range of masses, another dimension should   be added 
to the problem. This turns out to be   the initial density. 
Combining  the results of Carraro et al. (2001) with those
we are going to present here, the  monolithic collapse of baryonic
material may lead to different star formation histories ranging 
 from an initial prominent spike, to a sort of broad peak,
to a series of bursts of different intensity, and finally to an
ever continuing, 
fluctuating mode. The above trend holds   for a  given initial density at 
decreasing galaxy mass,
or for a given galaxy mass at decreasing initial density.

The layout of the paper is as follows.   Section~2 describes  
 the numerical tools and the set-up of the initial conditions of the 
galaxy models.
Section~3
presents the evolution  and  final properties of the galaxy models.
Section~4 deals with the analysis of the CMDs that based on the star
formation and chemical enrichment histories are expected to be observed at
the present time and compares them with those of  a few proto-type objects.  
Finally
Section~5 summarizes the results and draws some general conclusions.

\begin{figure}
\caption{The Color-Magnitude diagram 
 of two DSGs: on the left Carina, on the right Sculptor. This figure is 
taken from Smecker-Hane \& McWilliam (1999). }
\label{cmd_ca_scu}
\end{figure}

\begin{table}
\tabcolsep 0.2truecm 
\caption{Properties of some Dwarf Spheroidal Galaxies in the Local Group.
Data  from Mateo (1998).}
\begin{center}
\begin{tabular}{ccccc}
\noalign{\smallskip}\hline
\multicolumn{1}{c}{Galaxy}    &
\multicolumn{1}{c}{$M_{*}$  } &
\multicolumn{1}{c}{$r_e$}     &
\multicolumn{1}{c}{$\sigma$}  &
\multicolumn{1}{c}{$[Fe/H]$}   \\
& $10^{6} M_{\odot}$ & arcmin & km/sec &\\
\noalign{\smallskip}\hline
Carina     & 13 & 8.8$\pm$1.2  & 6.8$\pm$1.6 &  -2.0$\pm$0.2  \\
Phoenix    & 33 &              &             &  -1.9$\pm$0.1  \\
Leo~I      & 22 & 3.3$\pm$0.3  & 8.8$\pm$0.9 &  -1.5$\pm$0.4  \\
Sextans    & 19 & 3.2          & 6.6$\pm$0.7 &  -1.7$\pm$0.2  \\
Ursa Minor & 23 & 15.8$\pm$1.2 & 9.3$\pm$1.8 &  -2.2$\pm$0.1  \\
Draco      & 22 & 9.0$\pm$0.7  & 9.5$\pm$1.6 &  -2.0$\pm$0.15 \\
Sculptor   &  7 & 5.8$\pm$1.6  & 6.6$\pm$0.7 &  -1.8$\pm$0.1  \\
\noalign{\smallskip}\hline
\end{tabular}
\end{center}
\label{tab_data}
\end{table}

\section{N-Body, Tree-SPH Models }

The simulations presented here  have been performed using the Tree--SPH
code developed by Carraro at al. (1998) and Buonomo et al. (2000).
In this code,  the properties of the gaseous  component are
described by means of the Smoothed Particle
Hydrodynamics (SPH) technique (Lucy 1977 and Gingold \& Monaghan 1977), 
whereas the
gravitational forces are computed
by means of the hierarchical tree algorithm of Barnes \& Hut (1986), using
the tolerance parameter $\theta=0.8$,   expanding the tree nodes to quadrupole
order, and adopting a Plummer  softening parameter.

In the SPH method, each  particle represents a fluid element whose
position, velocity, energy, density etc. are followed in  time and space.
The properties of the fluid are locally estimated by an interpolation
which involves the smoothing length $h$.
In our code  each particle possesses its own time and space
varying smoothing length $h$, and evolves with its own time-step.
This renders the code highly adaptive and flexible, and
suited to speed-up  numerical calculations. No other details on the 
N-body Tree-SPH code in usage are given here for the sake of brevity. They 
can be found in Carraro et al (1998) and Buonomo et al (2000).

\subsection{Physical input}

{\it Star Formation.}   
The star formation rate (SFR) follows the  law  

\begin{equation}
\Psi(t) = \frac{d \rho_{\star}}{dt} = -\frac{d \rho_{g}}{dt} = -\frac{c_{\star} 
\rho_{g}}{t_{g}} 
\label{sf_rate}
\end{equation}

\noindent
where $c_{\star}$ and $t_g$ are the so-called  
dimensionless efficiency of star formation and  
the characteristic time scale, respectively. In these models we adopt
$c_{\star}=1$.
This characteristic time-scale $t_g$ is chosen to be the maximum between
the cooling,  $t_{cool}$, and the free-fall, $t_{ff}$, time scales 

\begin{equation}
t_{cool}={E \over |\dot E_H - \Lambda_C/\rho_g|}      \qquad\qquad   
t_{ff}= \Big( {1 \over 4 \pi G \rho} \Big)^{\frac{1}{2}} 
\end{equation}

\noindent
where $E$ is the current thermal energy per unit mass of gas, 
$\dot E_H$ is the heating rate per unit mass of gas by all possible sources
(supernova explosions, stellar winds, mechanical agents etc..),
and $\Lambda_C$ is the cooling rate (erg\,s$^{-1}$\,cm$^{-3}$) by radiative 
processes. For more details on 
$E_H$,  $\Lambda_C$ and associated cooling time scale see Chiosi et al. (1998). 
Finally, the free-fall time scale is calculated by means of the 
total density  $\rho$ (gas, stars, and DM).
Furthermore, a gas particle is eligible for star formation if the 
following conditions are met
(Katz 1992; Navarro \& White 1993):

\begin{description}
\item \centerline{ $t_{sound} > t_{ff}$ \qquad {\rm and} \qquad  
            $t_{cool} << t_{ff}$ }
\end{description}

\noindent
where $t_{sound}$ is the propagation time-scale of sound waves. 
It is worth recalling
that $t_{cool}$, $t_{ff}$ and $t_{sound}$ ultimately depend 
(although in a different fashion) 
on the density of the fluid.

{\it 
Radiative cooling.} The cooling rate  is taken from the  
tabulations of Sutherland \&
Dopita (1994) and Hollenbach \& McKee (1979) and amalgamated as in
Chiosi et al. (1998). 
This allows us to account for  the effects of variations
in the metallicity of the  fluid elements 
as a function of time and position.

{\it Energy Feed-Back.} 
This originates from  the SN\ae\ explosions of type Ia and II
(Greggio \& Renzini 1983) and   stellar winds from massive stars
(Chiosi \& Maeder 1986).  To clarify how the energy feed-back is 
implemented in the code, we briefly discuss how a gas particle is changed
into a star-particle. At any time step a gas particle eligible for star
formation decreases its mass by the quantity

\begin{equation}
\Delta m = m_g \Big[1 - exp( - {c* \Delta t \over t_g})\Big] 
\end{equation}
which simply follows from integrating equation (\ref{sf_rate}) 
over the time step,
and a new star-particle is created with mass $m_s= \Delta m$.
This in turn is conceived as being actually 
made of a large number of smaller sub-units (the real stars) 
lumped together and
distributed in mass according to a given initial mass function.
In other words a star-particle is a single stellar population whose
temporal evolution is well known (see for instance Bressan et al., 1994).
 For the purposes of the present
study we adopt the initial mass function of  Salpeter (1955) 
 over the mass interval from 0.1 to 120 $M_{\odot}$.  
As time elapses, the real stars in any star-particle evolve and die,
injecting gas (partially with the original composition and partially
enriched in metals) and energy by stellar winds and supernova explosions
(both of type II and type I). Since the IMF is known these quantities can be
easily calculated. Sharing of gas, metals and energy with 
neighbors is made according to the SPH formalism. While the new 
star-particle will now feel only the gravitational field, the left over 
gas-particle and all the others of the same type will continue to 
feel both gravitational and 
hydro-dynamical forces. The process can be repeated as long as there are 
gas-particles around eligible for star formation.

Two comments are  worthwhile here: 
(i) of the energy released to the ISM by a SN event (about
$\rm 10^{51} ergs$), only a tenth of it is given to the  inter-stellar medium
and the rest is radiated away (see Thornton et al. 1998
for an exhaustive discussion of this topic).
(ii) the energy injection by stellar winds may parallel that from 
SN\ae\ explosions.  
For all other details the reader should refer to the 
study by Chiosi et al (1998), whose prescriptions are adopted in the present 
models.

{\it Chemical enrichment}.  Supernova explosions and stellar winds 
increase the metallicity of the gas. The enrichment process 
 is described by means of the  closed-box model
 (Tinsley 1980; Portinari et al. 1998) applied to
each gas-particle of mass $m_g$. In brief, each
gas-particle is viewed as a  close box system, 
in which star formation occurs turning gas into stars and increasing
the metallicity according to the simple relation 
\begin{equation}
Z= y_Z\, \ln [{m_g\over m_g - m_s}]
\end{equation}
where $m_p$ is the initial mass of particle (gas), $m_s$ is the mass of 
stars borne in it, $y_Z$ is the so-called yield per stellar generation 
(constant as  long as the IMF is constant). We adopt here $y_Z=0.004$. 
See Tinsley (1980) and Portinari et al. (1998) for  definitions, more details 
and recent estimates of $y_Z$. 
Sharing of the metals among the gas-particles is 
described by means of a diffusive scheme (Groom 1997; Carraro et
al. 1998).
For more details on the above physical ingredients and their implementation in
the numerical code see Buonomo et al. (2000 and references therein).

\begin{table}
\tabcolsep 0.2truecm 
\caption{Initial conditions for the galaxy models.}
\begin{center}
\begin{tabular}{ccc cc}
\noalign{\smallskip}\hline
\multicolumn{1}{c}{Model}      &
\multicolumn{1}{c}{$M_{D}$  }    &
\multicolumn{1}{c}{$M_B$}      &
\multicolumn{1}{c}{$R_{D}$}      &
\multicolumn{1}{c}{$\rho_0$} \\
\noalign{\smallskip}
& $10^{9} M_{\odot}$ & $10^{9} M_{\odot}$ & kpc & $M_{\odot}/kpc^3$ \\
\noalign{\smallskip}\hline
A      & 0.9 & 0.1  & 4  & $3.9\times10^6$\\
\noalign{\smallskip}\hline
B1     & 0.9 & 0.1  & 16 & $5.8\times10^4$ \\
B2     & 0.9 & 0.1  & 26 & $1.3\times10^4$ \\
B3     & 0.9 & 0.1  & 35 & $5.5\times10^3$ \\
\noalign{\smallskip}\hline
\end{tabular}
\end{center}
\label{tab_ini}
\end{table}

\subsection{Initial conditions}

The initial conditions  of a galaxy are the total mass, $\rm M_T$,  made 
up in turn by DM  and baryonic matter (BM) with masses $M_{D}$  and 
$M_{B}$ respectively, the 
number of DM and BM particles,
their positions and velocities, and the mean density of the system
or equivalently its initial radius. All the simulations are made
using 10,000 particles for both  DM and BM. 

The following considerations are made to set up the initial conditions
listed  in Table \ref{tab_ini}.

(i) We start from the moment in which a lump of DM and BM (in standard 
cosmological  proportions) 
detaches itself  from the Hubble flow  and begins to collapse toward the
virial conditions. 
It is commonly assumed (see for instance Lacey \& Cole 1993, 1994 and 
references therein) that this occurs when the local
density $\rho(z)$ equals or exceeds the mean density of the Universe
$\rho_u(z)$ by a certain factor 

\begin{displaymath}
 \rho(z) = \langle \rho_0 \rangle \geq 200 \times \rho_u(z)
\end{displaymath}
where $z$ is the red-shift, $\langle \rho_0 \rangle$ is the initial
density of the proto-galaxy  and $\rho_u(z)$ is

\begin{equation}
\rho_{u}(z) = \frac{3 \,\, h^2 \times 100^2} {8 \pi G} (1+z)^3 \qquad\qquad{\rm or }
\label{rho_un}
\end{equation}
\begin{displaymath}
 \rho_{u}(z) = 1.99\times 10^{-29}\, h^2 (1+z)^3 \,\, \qquad {\rm g \,cm^{-3}}
\end{displaymath}
where $h=H_0/100$ is the normalized Hubble constant.
All other symbols have their usual meaning.

(ii)
Upon assigning the total mass $\rm M_T$ of the
proto-galaxy, we may  derive the 
initial radius,
 otherwise known as the virial radius or $R_{200}$,

\begin{displaymath}
R_{200}(z) = (\frac{3}{4 \pi})^{1/3} \times (\frac{M_{T}}{200
\rho_{u}})^{1/3}
\end{displaymath}

\begin{equation}
R_{200}(z) = 0.09617 \times 200^{-1/3} \times \Big({M_T\over h^2}\Big)^{1/3}
\times (1+z)^{-1}
\label{r_200}
\end{equation}

\noindent
where $M_T$ is in solar units, and $R_{200}$ is in kpc. 
In the following we will adopt $H_{0}$= 65 km/s/Mpc and $z_{for}=5$.

(iii) 
In order to simplify things we assume that the initial lump of DM 
very soon acquires 
 the condition of virial equilibrium, so that the distribution of DM
in the proto-galaxy has the  density  profile 

\begin{equation}
\rho(r) = \rho_c  {r_c \over  r}
\label{cent_sph}
\end{equation}

\noindent
where $r_c$ and $\rho_c$ are the radius and density, respectively, of
a small central sphere. 
The spatial coordinates of DM particles inside
 the initial sphere of radius $R_{200}$ 
are assigned by means of a {\it Monte-Carlo} procedure.

The initial velocity of a DM particle located at any radial distance
$r$  is derived from  the 
velocity dispersion $\sigma(r)$
for a spherical isotropic
collision-less system with the adopted density profile
(Binney \& Tremaine, 1987)

\begin{equation}
\rho(r)\sigma(r)^2 =  \int_r^{R_T} {G M(r') \over r'^2}\rho(r')dr'
\end{equation}
Inserting equation (\ref{cent_sph}) we get
\begin{equation}
\sigma(r)^2 =  \rho_c\, r_c \,G\, r\, \ln\left( \frac{R_T}{r}\right)   
\end{equation}

\noindent
where $R_T$ is the initial total radius, i.e. $R_{200}$. 
Finally, the velocity $v(r)$ is 
set equal to $v(r)= {1\over 3}\sigma(r)$ assuming equipartition among the three
components $v_x$, $v_y$, and $v_z$.

(iv) 
The BM inside  the DM halo follows the 
evolution of the latter, however,  no star formation is allowed to
 occur before  DM has   reached the virial condition.
The  particles of BM (gas at the beginning) 
 are distributed homogeneously inside the DM halo with
zero velocity field, thus mimicking the infall of primordial gas
into the
 potential well of DM (White \& Rees 1978).

(v)  The initial values for the mass of BM and DM are in the fixed proportion
 $M_{B}\simeq 0.1\, M_{D}$.
Since the numerical simulations are made using 
the same number of particles (10,000) for both DM and BM,
the mass resolution
of the DM is 10 times lower. The initial metallicity of the gas 
is assumed to be $Z = 10^{-4}$, the lowest value at our disposal in the grid 
of stellar models and related chemical yields.

(vi) 
The initial total mass 
of our models is chosen in such a way that the present day
total mass in stars agrees with the observational estimates for many
dwarf galaxies of the Local Group. According to Mateo (1998; see also the 
entries of Table~\ref{tab_data}) the mass in stars is about
10 to 30 $\times 10^{6} M_{\odot}$.  Basing on this we 
adopt the typical values $\rm M_T=1\times 10^{9} M_{\odot}$,  
$\rm M_D=0.9\times 10^{9} M_{\odot} $ and $M_B=0.1 \times 10^{9} M_{\odot}$. 
Owing to exploratory nature of this study the above choice is fully adequate 
to our purposes.

(vii) There is a final  consideration to be made concerning the
initial density (radii) of our models.
In principle, there should be one-to-one correspondence between
red-shift and over-density of the proto-galaxies with respect to the
surrounding medium. However, fluctuations around the mean excess
are always present. For the most massive galaxies the
degree of freedom is small because the fluctuations are small
at the large scales. This is not the case for galaxies of
lower mass because the fluctuations are larger at the smaller scales.
From a technical point of view we will consider two classes of
models: (a) a first group (labelled A) whose initial density strictly follows
relation (\ref{r_200}); (b) a second group (labelled B), in which the initial
density is arbitrarily set to the value 
$\rho_u(z)/200$ and let fluctuate around it.
Their initial radii
are accordingly expressed by relation (\ref{r_200}) in which the factor 
$200^{-1/3}$ is dropped.
It is worth stressing that this second group of models is only meant
to explore the consequences of varying the initial density of
the proto-galaxy. 
The analysis below will show that
the initial density
bears very much on the past SFH  
 passing from a single initial episode for high
densities to recurrent burst-like episodes for  low initial densities.

\begin{table*}
\caption{Properties of the Galaxy models.}
\begin{center}
\begin{tabular*}{130mm}{cc c c cc c c c} 
\noalign{\smallskip}\hline
\multicolumn{1}{c}{$Model$} &
\multicolumn{1}{c}{$M_{star}$} &
\multicolumn{1}{c}{$M_{gas}$} &
\multicolumn{1}{c}{$R_{e,B}$} &
\multicolumn{1}{c}{$\sigma_{B}$} &
\multicolumn{1}{c}{$\sigma_{D}$} &
\multicolumn{1}{c}{$\rho_{c,B}$} &
\multicolumn{1}{c}{$log(Z/Z_{\odot})_{Max}$}&
\multicolumn{1}{c}{ $log(Z/Z_{\odot})_{\Psi}$  }\\
\noalign{\smallskip}
  & $10^{7} M_{\odot}$ & $10^{7} M_{\odot}$ & Kpc & km/s & km/s & $M_{\odot}$/pc$^3$ \\
\noalign{\smallskip}\hline
\noalign{\smallskip}
A  & 3.2 & 6.8 & 0.07 & 10 &  25  &       &  -0.017 & -1.158 \\
\noalign{\smallskip}\hline
\noalign{\smallskip}
B1 & 7.7 & 2.3 & 0.24 & 4.7 & 18.7 & 0.70 &  -0.128 & -0.577 \\
B2 & 2.5 & 7.5 & 1.55 & 6.2 & 15.2 & 0.09 & -0.671 & -1.079 \\
B3 & 1.3 & 8.7 & 8.70 & 3.1 & 12.7 & 0.01 & -1.032 & -1.408 \\
\noalign{\smallskip}\hline
\end{tabular*}
\end{center}
\label{tab_res}
\end{table*}

\begin{figure}
\centerline{\psfig{file=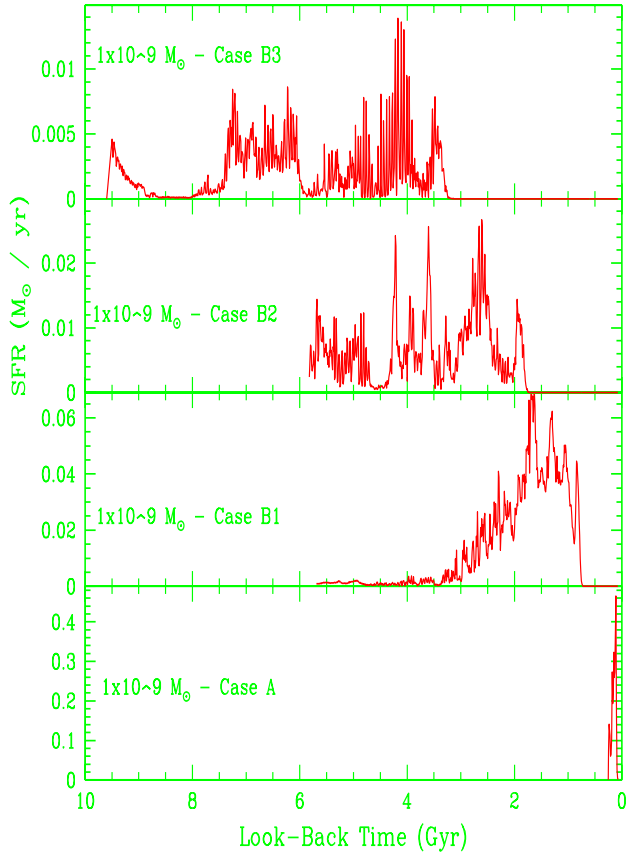,width=9cm,height=13cm}}
\caption{The SFR history in the four galaxy models with different mean 
initial density as a function of the look-back time in Gyr. The SFR is given
in $M_{\odot}$/year. The initial density
decreases from the bottom to the top panels (cf. 
Table \protect{\ref{tab_ini} }).}
\label{sfr_mod}
\end{figure}

\begin{figure}
\centerline{\psfig{file=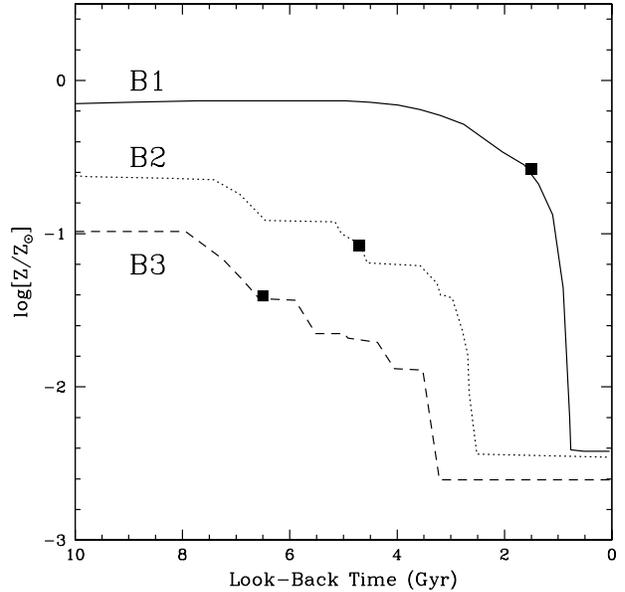,width=9cm,height=9cm}}
\caption{Metal enrichment of the three model galaxies of type B
as a function of the look-back time. The filled square along each curve shows 
the mean metallicity weighted on the star formation rate.}
\label{enrich}
\end{figure}

\section{The Model Galaxies}

The relevant data characterizing the  final state of evolution of the four
galaxies  
are summarized in Table \ref{tab_res}. These are the mass in stars, the 
mass of remaining gas, the present-day 
effective radius of the baryonic matter $R_{e,B}$, 
the central velocity dispersions of 
baryonic and dark matter, $\sigma_b$ and $\sigma_D$ respectively, and 
 an estimate of the maximum and mean metallicity
weighted on the star formation rate $\Psi(t)$,
 $\log (Z/Z_{\odot})_{Max}$ and $\langle \log (Z/Z_{\odot})\rangle_{\Psi}$,
respectively.
In the discussion below particular attention is given to models of type B
because they present the largest variety of SFHs. The results for the case
A galaxies are very similar to those of case B1.

\subsection{Star Formation and Metal Enrichment}

Inside the virialized halo of DM, gas cools and 
sinks toward the center of the gravitational potential well, 
and sooner or later forms stars according to the adopted 
prescription.

The four panels of  Fig.~\ref{sfr_mod}
show the SFR
 of our galaxies as a function of 
time. In  model A with the highest initial density 
the SFR reduces to a prominent,  initial
burst of activity followed by quiescence. All the action is within a 
narrow age range of about 0.2 Gyr. 

Passing to type B models the situation is more complex.  
Model B1 despite its   significantly
lower initial density compared to case A, still resembles this latter
because the star forming activity is confined in the past. However, two 
important new features appear: first star formation  is delayed for about
1 Gyr and second it stretches over an  age range of about 3 Gyr. The SFR is 
somewhat irregular and intermittent. Moving to even lower initial densities
(models B2 and B3) the features in question are amplified, the SFR shows 
indeed  many burst-like episodes of short duration (shorter than 0.1-0.2 Gyr) 
superimposed on a  gentle, nearly periodic trend 
 with typical time scale of about 1 Gyr.
 This behavior is natural in small size galaxies (Hirashita et al. 2000) 
 and 
is the consequence of the balance between cooling and heating as a function
of the depth of the potential well. More preciseley, the rate of star 
formation and heating in turn scale with the density, whereas the cooling rate 
scales 
with the square of the density. Therefore, in a high density environment
not only more stars are made and more energy and metals are injected but 
also cooling is more
efficient so that it easier for the gas to meet the star formation condition. 
Star formation is somehow forced to completion. In
contrast, in a low density medium, the rate of star formation and energy and
metals 
injection tend to be lower, but also cooling becomes   less efficient.
The gas particles find it more difficult to meet the conditions required 
for star formation. As a consequence of it, the star forming 
activity become fluctuating and even discontinuous.

In the course of evolution,  the  galaxies convert their
initial gas content into stars at different efficiencies. In general,
the higher the mean initial density, the larger is the fraction of gas
converted  into stars. However, this trend can be counteracted by the effects
of heating by supernova explosions, stellar winds etc. which  in the case
of very high efficiency such as in our case A may even lead to a lower fraction 
of gas being converted into stars (see the entries of Table~\ref{tab_res}
to be described below).

In all cases under consideration, 
the  gas left over by the star forming process 
is partly lost (out of the virial radius) and
partly locked in the outer DM halo. Finally the left-over gas is
partly enriched (30$\%$) and partly unprocessed (70$\%$) because of the partial
diffusive mixing.

In Fig.~\ref{enrich} we show the history of metal enrichment 
for the  models  B1, B2 and B3.  Model A is not displayed here because
thanks to the very initial spike of stellar activity and quiescence ever since,
the metallicity is practically constant with time.
In all other models we notice a continuous 
but irregular increase, characterized by some pauses and
several jumps, which mark successive bursts of star formation.
The maximum  metallicity goes from $\log(Z/Z_{\odot})_{Max}$=-1.032 (B3) to
$\log(Z/Z_{\odot})_{Max}$=-0.128 (B1) and $\log(Z/Z_{\odot})_{Max}$=-0.017 
for model A. However, the mean metallicity weighted on the star formation rate
$\langle \log(Z/Z_{\odot})\rangle_{\Psi}$ is much lower. It varies from
$\langle \log(Z/Z_{\odot})\rangle_{\Psi}$=-1.48 (B3) to 
$\langle \log(Z/Z_{\odot})\rangle_{\Psi}$=-0.577 (B1) and 
$\langle \log(Z/Z_{\odot})\rangle_{\Psi}$=-1.158 (A).

It is worth commenting here that these values of the metallicity are merely 
indicative of the real situation owing to our over-simplified treatment of
chemical enrichment.

\begin{figure}
\centerline{\psfig{file=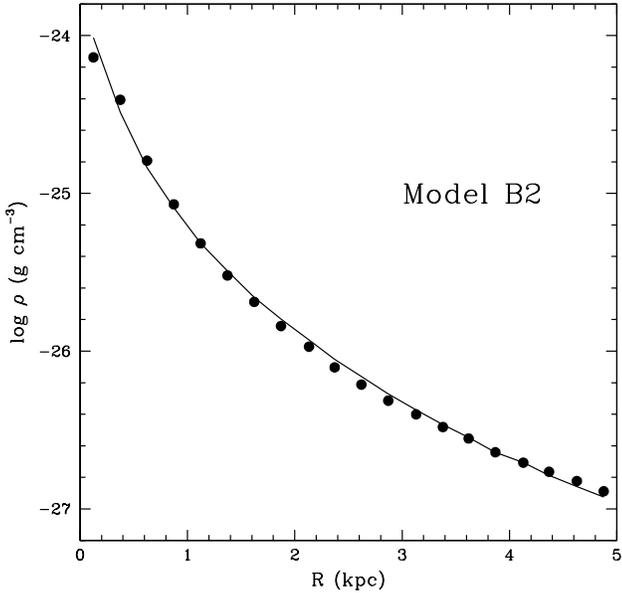,width=9cm,height=9cm}}
 \caption{The final  distribution of the star density (in g cm$^{-3}$) in model B2. 
Superimposed is a Hernquist profile
 for the effective radius $R_{e,B}$ = 1.55 kpc.}
\label{star_dis}
\end{figure}

\begin{figure}
\centerline{\psfig{file=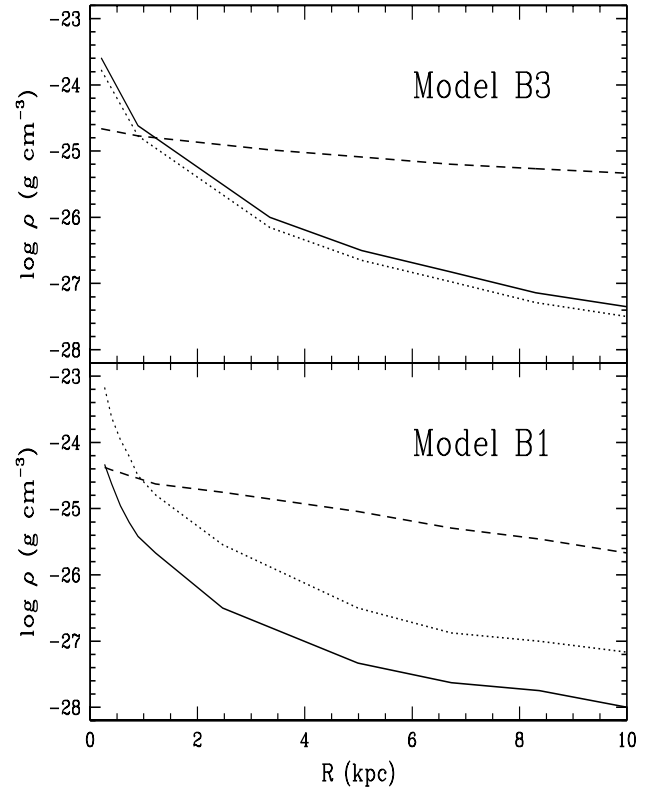,width=9cm,height=12cm}}
\caption{Final radial distribution of the gas (solid line), 
stars (dotted line) and DM (dashed line) densities (in g cm$^{-3}$) in the model
galaxies B1 and B3.}
\label{gas_stars_dm}
\end{figure}

\begin{figure}
\centerline{\psfig{file=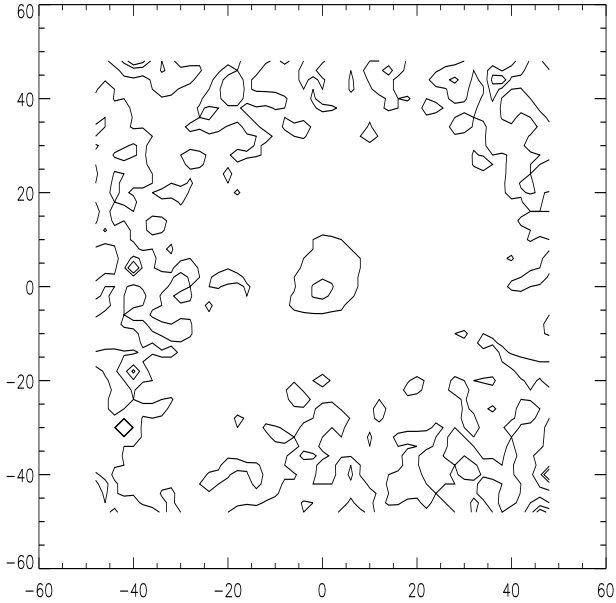,width=9cm,height=9cm}}
\caption{Iso-density contours of the gas distribution in  model B2 as seen
at the end of the simulation. The X and Y-axis are the spatial coordinates in
kpc  
centered on the center of the gravitational potential. 
Note the circular annulus void of gas surrounding the
central body of the galaxy. The gas in the outermost region is beyond the
virial radius and therefore escaping from the galaxy (galactic wind). }
\label{gas_dis}
\end{figure}

\begin{figure}
\centerline{\psfig{file=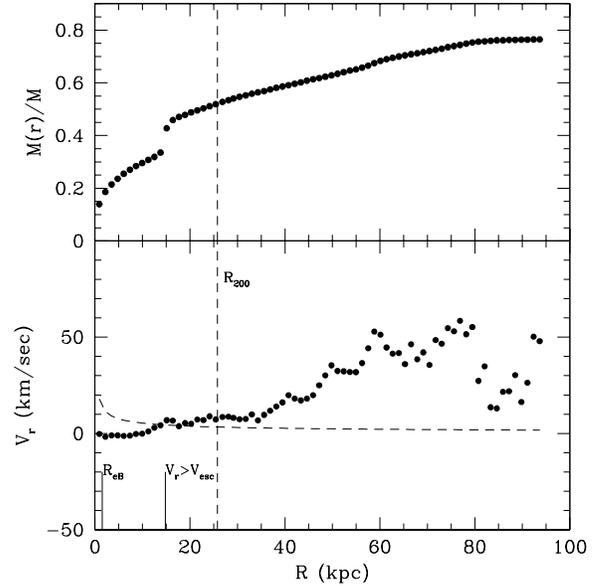,width=9cm,height=9cm}}
\caption {  Present-day gas and velocity profiles in the model galaxy B2. 
{\it Upper Panel}: The fractionary mass distribution
of gas as function of the radial distance in kpc. {\it Bottom panel}:
the radial velocity component $V_r(r)$ (dots) 
 and the escape velocity $V_{esc}(r)$ (long dashed line) 
as a function of the galacto-centric 
distance. The vertical dashed line is the virial radius, whereas the 
vertical solid lines show the effective radius of baryonic matter, $R_{e,B}$
and the layer, $r_{esc}$, at which $V_r(r) > V_{esc}$. All the gas located 
at $r > r_{esc}$ freely flows out.  }
\label{vel_gas_1e9nd}
\end{figure}

\subsection{Final structure of the model galaxies}

To illustrate how the mass of three main components,
i.e. DM, gas and stars, is distributed in our model galaxies
as a function of the radial distance, in  Fig. \ref{gas_stars_dm}  we show
the present-day radial profiles of the  mass density  in gas (solid line),  
stars (dotted line),
and DM (dashed line) limited to the cases of 
models B1 and B3.

\subsubsection{Dark Matter }

In our simulations, the DM halo surrounding the stellar and gaseous
matter does 
not  significantly change its properties in the course of evolution.
The density profile of DM is very smooth, and depending on the initial
conditions its central value ranges from a few $10^{-25}$ g/cm$^3$ 
to a few $10^{-23}$ g/cm$^3$.
The most important result is that the innermost regions of the
galaxies (within about 1 kpc of the center )  are dominated 
by baryons, whereas the outer parts
are dominated by DM (see Fig.~\ref{gas_stars_dm}). 
Moreover, the   velocity dispersion of the DM is about twice the 
velocity dispersion of the stars. 
These results agree with the  notion that
present-day DGs are  DM-dominated 
(Salucci \& Persic 1998), because  the original  content of baryonic
material has been largely lost in galactic winds (see below).

\subsubsection{Stars }

The final stellar mass of the models  is 
$3.2\times 10^7\, M_{\odot}$ for case A and it ranges from 
$1.3  \times 10^{7} M_{\odot}$ to about
$ 5.5 \times 10^{7} M_{\odot}$ going from B1 to B3.
The stellar component always dominates in the innermost regions of the 
galaxy models.
In order to check whether our models may resemble
real dwarf galaxies, the dwarf spheroidals in particular, 
we look closely  at the model B2. 
At the end of the simulation  the  central density of stars is  about 
$3 \times 10^{-24}$ g/cm$^{3}$ and the stellar matter occupies a central 
sphere of about 5 kpc radius (see Fig.~\ref{star_dis}),
 which is about five times
smaller than the initial virial radius, even though the density gradient is
so steep that most of the stellar 
mass is within the effective radius $R_{e,B}$.
The spherically averaged stellar density profile of the same model is 
then  compared with the Hernquist (1990) profile for 
the  stellar distribution in a dwarf spheroidal 
with effective radius $R_{e,B}$ = 1.55 kpc (see Fig.~\ref{star_dis}).
The remarkable agreement between the two profiles shows
that the final galaxy   resembles indeed  a dwarf spheroidal.
The effective radius of the model is about  17 times less than 
its initial  virial radius. 
Finally, we like to note that as for  kinematics
the stellar  velocity dispersion in the center 
of our model galaxies ranges from 10 to 3 km/s passing from
case A to cases B1, B2 and B3 (see Table~\ref{tab_res}).

\subsubsection{Gas }

Most of the initial gas originally inside the 
DM halo is not processed into
stars (see the entries of Table~\ref{tab_res}).

As expected in all models
but for minor differences the spatial distribution of the gas 
in the central region 
closely follows the distribution 
of the stellar content (see Fig.~\ref{gas_stars_dm}).
 
However, there is an interesting feature to note looking at the large scale
distribution of gas. The case is illustrated in Fig.~\ref{gas_dis}
for the proto-type model B2.
Of the gas left over   
inside the DM halo,  only the   fraction located in the central region
keeps
the original spatial distribution.
The remaining part,  heated up  by the energy feed-back from
supernova explosions and stellar winds,  reaches
regions very far from the central sphere containing most of the stars
(i.e. the luminous part of the galaxy).

Finally, it is worth noticing that  between the central part and the
most distant regions there is a very low density,  almost spherical shell,
which is due to the strong galactic wind episode occurred in the 
past.
This gas distribution  resembles indeed the distribution of
 cold gas recently
discovered by Blitz \& Robishaw (2000) in a sample of dwarf
spheroidal galaxies of the Local Group.

The major question is to be addressed whether or not part of the left over gas
becomes gravitationally unbound to the main body of the galaxy thus  freely 
escaping  as galactic wind.

\begin{table}
\tabcolsep 0.10truecm
\caption{Galactic winds.}
\begin{center}
\begin{tabular}{ccc ccc  c } 
\noalign{\smallskip}\hline
\multicolumn{1}{c}{$Model$} &
\multicolumn{1}{c}{$M_{g}$} &
\multicolumn{1}{c}{$R_{e,B}$} &
\multicolumn{1}{c}{$R_{200}$} &
\multicolumn{1}{c}{$M_{g,esc}$} &
\multicolumn{1}{c}{$\Delta M_{g,w}$}&
\multicolumn{1}{c}{ ${\Delta M_{g,w} \over M_g }$  }\\
\noalign{\smallskip}
  & $10^{7} M_{\odot}$ & 
     kpc & kpc&  $10^7 M_{\odot}$&$10^7 M_{\odot}$& \% \\
\noalign{\smallskip}\hline
A  &  6.8 & 0.07 & 4 &  2.4 & 4.5 &  66 \\
\noalign{\smallskip}\hline
\noalign{\smallskip}
B1 &  2.3 & 0.24 & 16 & 1.5 & 0.8 &  35 \\
B2 &  7.5 & 1.55 & 26 & 5.2 & 2.3 & 30 \\
B3 &  8.7 & 8.70 & 35 & 6.4 & 2.1 & 24 \\
\noalign{\smallskip}\hline
\end{tabular}
\end{center}
\label{tab_wind}
\end{table}

\subsubsection{Galactic winds }

The answer to the above question is given by  
 Fig.~\ref{vel_gas_1e9nd}, which displays the case of model B2,
 and Table~\ref{tab_wind}, which lists the relevant data for all the models.

The top panel of Fig.~\ref{vel_gas_1e9nd} shows the fractionary gas content
as a function of the radial distance in kpc, whereas the bottom panel
shows the radial 
component of the gas velocity $V_r(r)$ (dots) and the escape velocity 
$V_{esc}(r)$ (long dashed lines)
as a function of the galacto-centric distance.
The vertical, dashed line is the initial radius $R_{200}$, whereas 
the vertical solid lines labelled $R_{eB}$ and $V_r > V_{esc}$ show the
effective radius of BM and the layer at which $V_r > V_{esc}$.

There is a layer  $R_{esc}$ at which 
the condition $V_r > V_{esc}$ is met. All the gas lying above it can freely 
leave the galaxy in form of galactic wind. However, the amounts of gas lost
in the  wind varies with the type of model i.e. the initial density.
The situation is summarized in Table~\ref{tab_wind}
which lists the model identification label, the total among of left-over gas
$M_g$, the effective radius of BM $R_{e,B}$, the initial virial radius 
$R_{200}$, the gas content $M_{g,esc}$ up to at the layer at which the velocity
equals the escape velocity, the amount of gas lost in the wind 
$\Delta M_{g,w}$, and the percentage of gas lost in the wind with the respect to 
the total available gas, ${\Delta M_{g,w} \over M_g }$.

In model A, the gas mass blown away in the galactic wind amounts to about 66\%
of the remaining gas, whereas in models of type B it is roughly a factor of two 
less.

There is a final remark to be made: 
within the effective radius of BM, very little gas remains. This is 
nearly zero for model A, it amounts to $\simeq 5\%$ for model B1, 
to $\simeq 15\%$ for model B2,
and $\simeq 20\%$ for model B3. If our case A can be taken as representative 
of dwarf spheroidal galaxies, this would be consistent with the observed lack 
of gas in these systems (Mateo 1998). In contrast the reltively higher 
gas content in models B could make them more suited to interpret dwarf 
irregulars  (case B3) or dwarf spheroidals/irregulars (cases B1 and/or B2) 
that seem to be gas-rich (Mateo 1998),

\begin{table}
\tabcolsep 0.5truecm 
\caption{$T_G$ is the present age (in Gyr) of the  Galaxy models.
$T_{U,z{for}}$ is the age of the Universe at the epoch of galaxy
formation. $H_0$ is the Hubble constant in km/s/Mpc, whereas
$q_0$ is the deceleration parameter. Finally $z_{for}$ is the red-shift
at which galaxies are assumed  to form. }
\begin{tabular}{llrrr}
\hline
\multicolumn{1}{c}{$H_0$}      &
\multicolumn{1}{c}{$q_{0}$  }    &
\multicolumn{1}{c}{$z_{for}$}      &
\multicolumn{1}{c}{$T_{G}$}      &
\multicolumn{1}{c}{$T_{U,z_{for}}$} \\
\hline
50  & 0.  & 5   & 16.450 & 3.290  \\
50  & 0.5 & 5   & 12.265 & 0.895  \\
60  & 0.  & 5   & 13.708 & 2.742  \\
60  & 0.5 & 5   & 10.220 & 0.746  \\
70  & 0.  & 5   & 11.750 & 2.350  \\
70  & 0.5 & 5   & 8.760  & 0.640  \\
\hline
\end{tabular}
\label{tab_time}
\end{table}

\begin{figure}
\caption{A simulated CMD according to the SFR  
of case A presented in Fig. \protect{\ref{sfr_mod}}.
The mean metallicity is 
$\langle \log(Z/Z_{\odot})\rangle_{\Psi}$=-1.158.
The left panel is  for $H_0$=70, $q_0$=0.5,  $z_{for}$=5,
$T_G$=8.76 Gyr and  $T_{U,z_{for}}$=0.64 Gyr.
The right panel is the same but for $H_0$=50, 
$q_0$=0,  $z_{for}$=5, $T_G$=16.45  Gyr and $T_{U,z_{for}}$= 3.29 Gyr.
Finally, $1\sigma$ photometric errors are displayed at the right side of each
panel. 
}
\label{synt_cmd_A}
\end{figure}

\begin{figure}
\caption{The same as in \protect\ref{synt_cmd_A}, but for
case  B1. The mean metallicity is 
$\langle \log(Z/Z_{\odot})\rangle_{\Psi}$=-0.577.
The left panel is  for $H_0$=70, $q_0$=0.5,  $z_{for}$=5,
$T_G$=8.76 Gyr and  $T_{U,z_{for}}$=0.64 Gyr.
The right panel is the same but for $H_0$=50, 
$q_0$=0,  $z_{for}$=5, $T_G$=16.45  Gyr and $T_{U_{for}}$= 3.29 Gyr. }
\label{synt_cmd_B1}
\end{figure}

\begin{figure}
\caption{The same as in \protect\ref{synt_cmd_A}, but for
case B2. 
The mean metallicity is $\langle \log(Z/Z_{\odot})\rangle_{\Psi}$=-1.079.
The left panel is  for $H_0$=70, 
$q_0$=0.5,  $z_{for}$=5, $T_G$=8.76 Gyr and 
 $T_{U,z_{for}}$=0.64 Gyr.
The right panel is the same but for $H_0$=50, 
$q_0$=0,  $z_{for}$=5, $T_G$=16.45  Gyr and 
$T_{U,z_{for}}$= 3.29 Gyr. }
\label{synt_cmd_B2}
\end{figure}

\begin{figure}

\caption{The same as in \protect\ref{synt_cmd_A}, but for 
case B3. The mean metallicity is 
$\langle \log(Z/Z_{\odot})\rangle_{\Psi}$=-1.408.
The left panel is  for $H_0$=70, 
$q_0$=0.5,  $z_{for}$=5, $T_G$=8.76 Gyr and 
is $T_{U,z_{for}}$=0.64 Gyr.
The right panel is the same but for $H_0$=50, 
$q_0$=0,  $z_{for}$=5, $T_G$=16.45  Gyr and 
$T_{U,z_{for}}$= 3.29 Gyr. }
\label{synt_cmd_B3}
\end{figure}

\section{Color-Magnitude Diagrams}

The aim of this section is to check whether the SFR and 
enrichment law of our model galaxies
may generate CMDs like those  observed today in
dwarf galaxies of the Local Group.

Prior to this, one has to choose the link 
between  the red-shift of galaxy formation, $z_{for}$ and age
in order to convert the rest-frame ages used  so far into 
absolute ages of the stars which would now be observed. 
This means that a certain model for
the Universe has to be adopted.
For the purposes of the present study we adopt the Friedman model
described by the Hubble constant $H_0$ and deceleration parameter $q_0$.
In all simulations we assume the same red-shift of galaxy formation,
i.e. $z_{for}=5$.
To facilitate the use of the simulations below, in Table \ref{tab_time} we 
summarize the maximum  ages $T_G$ of the galaxy models, and the age 
$T_{U,z_{for}}$ of the Universe  
at $z_{for}=5$, for three different values
of $H_0$ (50, 60 and 70 km/s/Mpc), and two values of $q_0$
(0 and 0.5). Higher values  of both quantities are not
likely here because they would predict galactic ages  $T_G$ that are too
old, at least
compared to what is currently known about the age of globular clusters, i.e.
from 10.0 to 15.8 Gyr as recently reviewed by 
Carretta et al. (2000).  

Synthetic CMDs
are constructed by means of a population
synthesis algorithm. Given the galaxy age $T_G$, for each value of 
the stellar age $t$, an isochrone with 
the mean metallicity  $\langle Z \rangle$
is derived from a set of stellar evolutionary tracks.
These are from Girardi et al.\ (2000) for $0.0004\le Z\le 0.03$, 
and  Girardi et al.\ (1996)  for $Z=0.0001$. 
The evolutionary sequences  span the mass and age ranges of 
$0.15 \, M_{\odot} \le M \le 7\, M_{\odot}$, and from 0.06 to 20 Gyr.
The mass and age ranges are
fully adequate to describe the stellar populations of dwarf spheroidal 
galaxies. All  important evolutionary phases 
are included, from the zero-age main 
sequence to the end of the thermally-pulsing asymptotic giant branch.
The
isochrones in the theoretical plane (luminosity versus 
 effective temperature)
are transformed to the observational 
magnitudes and colors using the  
transformations provided by Bertelli et al.\ (1994).

Each isochrone of age $t$ is  populated by stars 
according to the Salpeter initial
mass function (IMF), and
in proportions fixed by  the star formation rate, SFR$(T_G-t)$. 
The  relative number of stars predicted in each section of the CMD 
is added 
to  the composite population. The procedure 
continues for all  values of the age, i.e. 
from the present ($t=0$)
up to the  maximum galaxy age ($t=T_G$).

Finally, the CMD is  populated using a arbitrary total number 
of stars. In the 
present simulations, we  include the effects of photometric 
errors and neglect those  of binaries for the sake of
simplicity. Photometric errors are assumed to be similar to those
obtained by Smecker-Hane \& McWilliam (1999) in HST observations of Carina 
dSph.

The most uncertain ingredient  of these CMD simulations is
mass loss along the red giant 
branch (RGB) of low-mass stars. This 
is taken into account by 
decreasing the stellar mass when passing from the tip of the
RGB to the zero-age horizontal branch (see Renzini 1977). 
Mass loss is assumed to follow the Reimers (1975) law with a suitable 
value of the parameter  $\eta$ (Fusi-Pecci \& Renzini 1976).
It is worth recalling that by  applying mass loss at the stage 
of isochrone 
construction (and not during the computation of evolutionary 
tracks) allows us to easily vary $\eta$ from
 simulation to simulation if required. 
We adopt  the  mean value 
$\langle\eta\rangle=0.50$ with dispersion $\sigma_{\eta}=0.05$. These values
are slightly larger that 
those used to reproduce the morphology of globular clusters
(Renzini \& Fusi-Pecci (1988) but still within current uncertainties.

The  above technique is used to generate the CMDs shown
in Figs. \ref{synt_cmd_A} through    \ref{synt_cmd_B3}
for the  SFHs (Fig.~\ref{sfr_mod}) and the enrichment rate (Fig~\ref{enrich})
of our model galaxies A,
B1, B2, and B3. 
Furthermore,  CMDs are presented for  two, somewhat extreme values of 
the galaxy age $T_G$ corresponding to   different values 
of $H_0$,   $q_0$,  $z_{for}$,  $T_{U,z{for}}$  as indicated.

{\it Model A.}
Case A is straightforward because owing to its prominent and
unique initial burst of activity, the associated CMD would simply resemble 
that of a globular cluster. This is shown in the two panels of
Fig.\ref{synt_cmd_A}. The left panel is for $H_0$=70, 
$q_0$=0.5,  $z_{for}$=5 and $T_G$=8.76 Gyr, whereas the right panel is
for $H_0$=50, 
$q_0$=0,   $z_{for}$=5, and $T_G$=16.45 Gyr. The two panels  
clearly show how the CMD
of  a {\it globular cluster } with intermediate mean  metallicity 
($\langle\log Z/Z_{\odot}\rangle_{\Psi}$=-1.158) changes
as a function of the age. Passing from the young (left) to the old (right)
population, the turn-off luminosity decreases whereas the clump of red
stars (He-burners) turns into an extended 
horizontal branch. This is more populated at the red side because  the 
 metallicity is somewhat higher than the typical value in globular clusters
 with very extended horizontal branches, mostly populated at the blue side. 

{\it Model B1.} The left and right panels of  Fig.\ref{synt_cmd_B1} 
show the expected CMD for the models B1.
The CMD of the  older age (right panel)
much resembles the one of an old globular cluster with relatively high
metallicity, 
$\langle\log Z/Z_{\odot}\rangle_{\Psi}$=-0.577 according
to the entries of
Table \ref{tab_res}.  There is some blurring at the turn-off and 
sub-giant branch due to the large age spread caused by
the prolonged initial  stellar activity (about 3 Gyr). These effect adds a 
new dimension to the horizontal branch, because it is now populated by stars
with significantly different initial mass, which tends to evenly populate
the horizontal branch. Despite the high metallicity, this CMD
could closely mimic the ones of 
Sculptor (see Fig.~\ref{cmd_ca_scu}) 
or Ursa Minor and Draco,  i.e.
galaxies dominated by a single burst of SF which extended for some time
(see Hernandez et al. 2000).
By decreasing the age down to 9 Gyr, the red clump 
(typical of an intermediate age population) appears. Although the star 
formation 
lasted for about 3 Gyr, its ever decreasing efficiency past the initial 
peak does not allow the presence of a secondary distinct generation of stars
but only a large scatter at  the turn-off. This CMD is a good
candidate to explain the properties of
 Leo~II (Hernandez et al. 2000).

{\it Model B2.}
The panels of  Fig.\ref{synt_cmd_B2} 
show the expected CMD for the same value of $z_{for}$, 
$H_0$ and $q_0$=0 as in Figs.~\ref{synt_cmd_A} and \ref{synt_cmd_B1} 
but for model B2.
The CMD of the right panel does not significantly differ from its analog 
in Fig.~\ref{synt_cmd_B1} but for a larger scatter at the turn-off, 
sub-giant branch,  and horizontal branch.  
The CMD  in the left panel owing to its much younger 
age clearly shows the several distinct bursts of star formation already 
noticed in Fig.~\ref{sfr_mod}.

{\it Model B3.}
Finally we have the case of models B3, whose CMD is shown in the left and 
right panels of 
\ref{synt_cmd_B3} for the same combinations of $H_0$, $q_0$ and $z_{for}$
as in the previous simulations.
The results of the right panel 
($H_0$=50, $q_0$=0, $T_G=16.45$ Gyr) are
particularly interesting because three distinct old populations are now
visible going from old to relatively young (say about 6 Gyr).
Galaxies like Leo~I or Carina could be   reproduced by a star formation
history like this.
 See for instance the CMD of 
Fig.~\ref{cmd_ca_scu} for Carina, or that of  Leo\,I given by
 Held et al. (2000). 
The CMD of the left panel, owing to the much younger ages involved and almost
continuous star formation up to the present
could perhaps mimic  the case of a DIrr.

{\it A general remark.} The above simulations of dwarf galaxies
and associated CMDs were not designed to match any particular
dwarf galaxy. In most cases, the metallicity of the models was
too high compared to observational value. This is certainly due to
the adoption of the closed-box description. Work is in progress 
to improve upon the treatment of chemical enrichment. Our
simulations  are  intended only to show that dwarf galaxies with the same
total mass
of $10^{9} M_{\odot}$ but originated from proto-galaxies of different 
initial densities may undergo different
histories of star formation which in turn yield different
CMDs, in agreement with the large variety of observational
data for such  galaxies. Dwarf galaxies formed in a high 
density environment tend to have a single initial burst of stellar 
activity. In contrast,  galaxies of the same mass but with low initial
density tend to undergo star formation over long periods of
time, showing a number of burst-like episodes. In some extreme
cases star formation may stretch up to  the present. 
Accordingly we pass from galaxies that would nowadays 
appear as DSph, DE and DIrr.

\section{Summary and conclusions}

We have tried to understand the physical reason for the large variety of star
formation histories indicated by the stellar content of dwarf galaxies,
with particular attention to the case of dwarf spheroidals and ellipticals.
Indeed objects whose estimated mass in stars, mean metallicity, and velocity
dispersion span a rather narrow range of values, can have very different
star formation histories.  

With the aid of Tree-SPH, N-Body simulations of galaxy formation and evolution
we find that  systems of the same total (dark and baryonic) mass but
different initial densities may have different star formation histories
ranging from a single initial episode to prolonged stellar activity
extended  into a series of bursts over a large fraction of the Hubble time.

Starting from "realistic" initial conditions, we follow the monolithic
collapse of baryons inside  the virialized halo of DM, and the formation 
of stars by gas cooling and collapse. The models include accurate treatments of
star formation, chemical enrichment, energy feed-back from supernova explosions
and stellar winds, and metal-dependent cooling.

The structural properties of the 
 models remarkably agree with the observational data 
of real galaxies (mass density profiles, central velocity dispersions, mean
metallicities, mass-to light ratios, and even gas distributions). 

Furthermore,  
using the theoretical star formation and metal enrichment histories
of the model galaxies
we simulate the  CMDs that one would expect to observe today
as a function of the underlying cosmological parameters $H_0$, $q_0$,
$z_{for}$ and $T_G$. In many cases the simulated CMDs remarkably resemble
 the observational ones.

The main conclusion of this study is that the adopted scheme of galaxy
formation, i.e. early virialization of DM haloes and monolithic collapse of
baryons into their potential wells, may lead to a variety of different
situations as far as the detailed star formation histories is concerned,  that
depend on the initial density of the system.

Together with the results of the companion paper by Carraro et al. (2001), 
briefly summarized in Section 2,  this study suggests that 
total mass and initial density are the key parameters
governing the final stellar content of  elliptical galaxies. Ample
possibilities  exist for star formation histories, ranging from
a single initial episode of short duration, to large time intervals, to 
burst-like activities, and   even to very prolonged  star formation.

A final consideration is worth being made here. Throughout the analysis,
when referring to the initial density this has been linked to red-shift via
the over-density with respect to the mean density of the Universe 
required by a proto-galaxy to detach himself from the Hubble flow 
and to collapse.
This indeed was the over-density assigned to model A. Strictly speaking,
at the red-shift of galaxy formation we have considered only models of type A
are possible. To have models of type B brought into existence one has to
wait for a fall-off of the over-density by a factor of 70 at least, or 
equivalently $\Delta z_{for}\simeq 3$. In other words, 
galaxies of similar mass but formed at $z_{for}\simeq 2$ would behave as 
models B and
for $50 \leq H_0 \leq  60$ and $q_0\simeq 0$ would have a maximum age for
their stellar populations still fully compatible with the information
derived from the observational CMDs.
In other words, proto-galaxies of the same total mass collapsing at different
red-shifts would contain different kinds of stellar populations.

\section*{Acknowledgments}

This study has been financed by the Italian Ministry of University, Scientific
Research and Technology (MURST) under contract {\it Formation and Evolution of
Galaxies} n. 9802192401.

\label{lastpage}

\end{document}